\documentclass[twoside,11pt]{article}
\def\be{\begin{equation}}
\def\ee{\end{equation}}
\def\ba{\begin{align}}
\def\ea{\end{align}}
\def\bea{\begin{eqnarray}}
\def\eea{\end{eqnarray}}

\def\n{\nu}

\def\ba{\bar{\mathcal A}}

\usepackage{amsmath,amssymb,amsfonts}
\usepackage{amsfonts}
\usepackage{graphicx}
\usepackage{graphics}
\usepackage{breqn}
\begin{document}
\setlength{\unitlength}{10mm}

\title{Gauging the Relativistic Particle Model on the Noncommutative plane}

\author{
Salman Abarghouei Nejad
\\
{\scriptsize Department of Physics,
}\\{\scriptsize University of Kashan, Kashan 87317-51167, I. R. Iran}
\\
\\
Mehdi Dehghani
\\
{\scriptsize Department of Physics, Faculty of Science,
}\\{\scriptsize Shahrekord University, Shahrekord, P. O. Box 115, I. R. Iran}
\\
\\
Majid Monemzadeh
\\
{\scriptsize Department of Physics,
}\\{\scriptsize University of Kashan, Kashan 87317-51167, I. R. Iran}}

\date{}
\maketitle
\vspace{1cm}
\vspace{-1.5cm}\begin{abstract}
\noindent  We construct a new model for relativistic particle on the noncommutative surface in $(2+1)$ dimensions, using the symplectic formalism of constrained systems and embedding the model on an extended phase space. We suggest a short cut to construct the gauged Lagrangian, using the Poisson algebra of constraints, without calculating the whole procedure of symplectic formalism. We also propose an approach for the systems, in which the symplectic formalism is not applicable, due to truncation of secondary constraints appearing at the first level. After gauging the model, we obtained generators of gauge transformations of the model. Finally, by extracting the corresponding Poisson structure of all constraints, we show the effect of gauging on the canonical structure of the phase spaces of both primary and gauged models. 



\end{abstract}

\newpage

\newcommand{\f}{\frac}

\newtheorem{theorem}{Theorem}[section]
\newcommand{\sta}{\stackrel}
\section*{Introduction}
Noncommutativity can be added to spatial coordinates' properties naturally, using Dirac's approach for constrained systems in the framework of quantum field theory \cite{3,d12}. This approach leads us to construct non-relativistic theories in noncommutative (NC) space. 

It has been shown that adding a Chern-Simons type term for the coordinate type variable, $v_{\mu}$, to the first order action of a relativistic commutative system, one can obtain a relativistic NC theory \cite{d}.
\begin{equation}
S_{NC}=S_{C}(q_{\mu},v_{\mu})+\int  \dot{v}_{\mu} \theta^{\mu\nu}v_{\nu} \textit{d}\tau,
\end{equation}
 Thus, we obtain a NC theory which intrinsically owns a Poincare structure. It means that NC variables' brackets are invariant under the Poincare transformations \cite{d7}. So, the proper action will be,
\begin{equation}
S=\int \textit{d}\tau [\dot{x}^{\mu}v_{\mu}-\frac{e}{2}(v^{\mu}v_{\mu}-m^{2})+\dot{v}_{\mu}\theta^{\mu\nu}v_{\nu}].
\end{equation}
The first term in the above action makes it a first order action. Also, the second term impose the relativistic dispersion relation as a constraint into the action. In addition, the last term transforms the spatial plane on which the particle live, to a NC plane.
which $e$ is the einbein variable in the action, which is added in order to include the massless cases. 

It goes without saying that adding the Chern-Simons type term to the first order action of the relativistic particle, imposes the relativistic mass shell constraint, $p_{\mu}p^{\mu}-m^{2}$ to the action \cite{d}. This constraint is a first class one, according to Dirac's classification of constraints \cite{3}. He, as the pioneer of constrained systems, tried to solve the quantization problems of any system by presenting the theory of constrained systems, and classifying constraints into primary, secondary, first and second class ones \cite{3}. He also expressed that first class constraints are the generators of gauge transformations \cite{dirac2}.

But, the use of the  Dirac's quantization method needs a special care, because there may be some nonphysical degrees of freedom in the system, which spoils such an approach. The advent of these nonphysical degrees of freedom are due to the presence of another kind of constraints, called second class ones \cite{3}.

Hence, in order to gauge a system, these nonphysical degrees of freedom must be eliminated by converting second class constraints to first class ones. In order to do that there has been existed some approaches such as BFT \cite{Monem1,BFFT1,BFFT2,BFFT3,9}, and FJ formalisms \cite{1,8,2}. The main procedure of these methods is based on the embedding process. This procedure is based on this fact that the noninvariant system will be embedded in an extended phase space in order to change the second class nature of constraints into first class ones. It is very important to know that the equations of motion of the original system and the final system obtained via embedding method are equivalent. Also, these methods divulges all hidden symmetries of the original system. 

The BFT approach to gauge the NC relativistic particle has been studied before \cite{dehghani}. In this article we gauge the mentioned system on the two dimensional spatial NC plane, i.e.$ (\theta^{0\mu}=0) $, using the symplectic formalism, which is a newly updated version of FJ method \cite{8} in order to investigate the effect of embedding on the NC plane. This formalism which is based mathematically on the symplectic structure of the phase space, is existed to keep us away from the consistency problems which ruins Poisson brackets algebra and consequently spoils any quantization technique in constrained systems \cite{2}. 

Unlike its congeners, in the symplectic formalism, all constraints are assumed to be equivalent and so, there will be no distinction between first and second class constraints \cite{xu}. So, an important feature of this formalism is its difference from other gauging approaches \cite{4,5}.

To use symplectic formalism, we should start using the first order Lagrangian, which its corresponding equation of motion does not imbue any acceleration. This is necessary to obtain Hamiltonian equation of motion from the variational principle \cite{2}. Also, any second-order Lagrangian should be transformed into first order one by expanding the configuration space, including conjugate momentum and coordinate variables \cite{6}. Then, in order to enlarge the phase space and start embedding process, a new variables has been introduced to linearize the system, which is named, as the Wess-Zumino (WZ) variables \cite{1}. At the end, one can use the Legendre transform to pass from Lagrangian to Hamiltonian \cite{7}.

In current article, we introduce the symplectic Lagrangian in the first section. Then, in the second part the symplectic formalism is presented and we show that this method may be failed for some particular cases, such as the relativistic particle on the NC surface and we present the proper solution by adding some auxiliary fields to the model and eliminating them at the end of the gauging process. In the third section, we apply the symplectic formalism on the gauge fixed model, and finally, in the last section we purify the constraints and quantize the model.

\section{Gauged Lagrangian}
A gauged Lagrangian of a typical system without gauge symmetry, $ \tilde{L}^{(1)} $, will be obtained by adding a Lagrangian depending on the new dynamical variable called WZ variable to the first order Lagrangian,
\begin{equation}
\tilde{L}^{(1)}= L^{(1)}+ L_{WZ},
\end{equation}
where, $ L_{WZ} $ is a function depending on the original coordinates and WZ variable. 

In order to obtain this added term, there has been driven an iterative differential equation with the help of symplectic two form zero modes and the potential of the model \cite{8}. As a matter of fact, for most cases, and particularly for this studied model, the mentioned iterative procedure will not continue more than two levels. Thus we introduce a short cut formula to make the WZ Lagrangian. 

Let's imagine that such models has some primary constraints which are denoted by $ \phi_{i} $. To start with, we should find the constraint which is first class in comparison to other primary constraints. According to Dirac's guess, the presence of the first class constraint guarantees the presence of a gauge symmetry in the model. We call this primary first class constraint as $ \bar \phi_{j} $,
\begin{equation}
\{\phi_{i},\bar \phi_{j} \}=0.
\end{equation}
By applying the symplectic approach, we will obtain the secondary constraint, denoted by $ \phi_{i}^\prime $.
We construct the WZ Lagrangian by adding two generators $ G^{(1)} $ and $ G^{(2)} $, as,
\begin{equation}
L_{WZ}=G^{(1)}+G^{(2)},
\end{equation}
which are defined by following relations,
\begin{align}\label{Condition}
& G^{(1)}=\sigma \phi_{i}^\prime \quad,  \nonumber \\
& G^{(2)}=-\sigma^{2}\{\phi_{i}^\prime,\bar\phi_{j} \} .
\end{align}
In the above equation, $ \sigma $ is the WZ variable, and its conjugate momentum, $ p_\sigma $, which will not be appeared in the gauged model is a first class constraint. Thus, it is the sign of the presence of the first class constraint in the obtained model.
\section{Symplectic Formalism}
To start with, we use the first iterated Lagrangian, $ L^{(0)} $,as
\begin{equation}
L^{(0)}=\dot{x}_{\mu} v^{\mu} -\frac{1}{2}e(v_{\mu} v^{\mu}-m^{2})+\dot{v}_{\mu} \theta^{\mu \nu} v_{\nu},
\end{equation}
with which we can read off the canonical Hamiltonian as,
\begin{align}\label{0H1}
& H_{c}= \dot{x}_{\mu}p^{\mu}+\dot{v}_{\mu}\wp^{\mu} -L^{(0)} , \nonumber \\
& \quad = \frac{1}{2}e(v_{\mu} v^{\mu}-m^{2})=V^{(0)}.
\end{align}

Obtaining the corresponding constraints,
\begin{align}\label{cons12}
& \phi_{1\mu}=p_{\mu}-v_{\mu}, \nonumber \\
& \phi_{2\mu}=\wp_{\mu}-\theta_{\mu \nu} v^{\nu}.
\end{align}
In order to construct the first order Lagrangian from the free particle Lagrangian, we add constraints to that, using new dynamical variables as  undetermined Lagrange multipliers, $\lambda_{i}$.
\begin{equation}\label{L0bar}
    L^{(0)}= \dot{x}_{\mu}P^{\mu}+\dot{v}_{\mu}\wp^{\mu} -H_c-\lambda_{i}\phi^{i}.
\end{equation}

In order to remove the constraint from Hamiltonian and add it to the kinetic part of the Lagrangian, we substitute \eqref{cons12} in the Lagrangian \eqref{L0bar}. As a result, the first iterative Lagrangian will be obtained as,
\begin{equation}
\frac{1}{2} (v_{\mu} v^{\mu}-m^{2}) -\sum_{i=1}^{2}\dot{\lambda}_{i}\phi^{i} = \dot{x}_{\mu}p^{\mu}+\dot{v}_{\mu}\wp^{\mu} -L^{(1)}.
\end{equation}
Thus we have,
\begin{equation}\label{0L(1)}
L^{(1)}=\dot{x}_{\mu}p^{\mu}+\dot{v}_{\mu}\wp^{\mu} -V^{(1)}+\dot{\lambda}_{1\mu}(p_{\mu}-v_{\mu})+\dot{\lambda}_{2\mu}(\wp_{\mu}-\theta_{\mu \nu} v^{\nu}),
\end{equation}
in which,
\begin{equation}\label{0V(1)}
V^{(1)}=\frac{1}{2}e(v_{\mu} v^{\mu}-m^{2}).
\end{equation}

The first iterated symplectic variables and corresponding conjugate momenta are now defined as,
\begin{align}
& \xi_{\alpha}^{(1)}=(x_{\mu},v_{\mu},p_{\mu},\wp_{\mu},\lambda_{1\mu},\lambda_{2\mu}),  \nonumber \\
& \mathcal A_{\beta}^{(1)}=(P_{\nu},\wp_{\nu},\textbf{0}_{\nu},\textbf{0}_{\nu },\phi_{1\nu},\phi_{2\nu}),
\end{align}
\begin{description}
\item[Note;]
 In this article, the Greek indices, $ \alpha $ , $ \beta $, $ \bar \alpha $, $ \bar \beta $, $ \tilde \alpha $, and $\tilde \beta $, are used to determine the phase space variables, while, $ \mu $ and $ \nu $, $ (\mu,\nu=0,1,2) $, indicate the dimensions of the space-time. Also, $ \textbf{0}_{\mu} $ (or simply $ \textbf{0} $) is a vector that all its components are zero, and $ 0 $ is a scalar. Moreover, the number of constraints are  counted with the help of indices $ i $ and $ j $.
\end{description}
Thus, here, $ \alpha $ and $ \beta $ are counted form $ 1 $ to $ 18 $.

Using $ f_{\alpha\beta}=\partial_{\alpha} \mathcal A_{\beta}-\partial_{\beta} \mathcal A_{\alpha} $, we obtain the corresponding symplectic two form, $f_{\alpha \beta}^{(1)}$.
\begin{eqnarray}
 f^{(1)}_{\alpha \beta}=\left( \begin{array}{cccccc}
  \textbf{0} & \textbf{0} &  -\delta _{\nu }^{\mu } &  \textbf{0} &  \textbf{0} &  \textbf{0} \\
  \textbf{0} &  \textbf{0}  & \textbf{0} & -\delta _{\nu }^{\mu } &  -\delta _{\nu }^{\mu } & -\theta ^{\mu \rho }  \\
  \delta _{\mu }^{\nu } &   \textbf{0}  &   \textbf{0} &  \textbf{0} & \delta _{\nu }^{\mu } &   \textbf{0}   \\
   \textbf{0}  & \delta _{\mu }^{\nu }  &   \textbf{0} &  \textbf{0}&   \textbf{0}   & \delta _{\nu }^{\mu }  \\
 \textbf{0} & \delta _{\mu }^{\nu }  & -\delta _{\nu }^{\mu } & \textbf{0}  & \textbf{0} & \textbf{0}   \\
 \textbf{0} & \theta ^{\mu \rho } & \textbf{0} & -\delta _{\nu }^{\mu } & \textbf{0} & \textbf{0}    \\
 \end{array}
\right).
\end{eqnarray}
Apparently, this tensor does not have any zero mode. Hence, it does not generate any extra secondary constraint.

Considering relations \eqref{Condition}, we will see that no WZ term will be added to the first order Lagrangian \eqref{0L(1)}. Thus, all attempts to gauge this model fails in the usual manner of symplectic gauging.

In order to overcome this problem, we can add some auxiliary fields to the first order Lagrangian which converts the first class constraints to second class ones and then the symplectic method can be applied to the system \cite{9}. 

The following extensions convert first class constraints to second class ones,
\begin{equation}
p_i \rightarrow p_i+\eta_i,
\end{equation}
where $ \eta_i $ and $ p_i $ are auxiliary conjugate variables, while this transformation in the Lagrangian formalism can
be done by the following replacement,
\begin{equation}
L \rightarrow L-\eta_{i} \dot{q}_{i}+\frac{1}{2}\sum_{i} \dot{\eta}_{i}^{2}.
\end{equation}
This replacement is a gauge fixing term, inserted in the gauge invariant Lagrangian. Though, the new Lagrangian gives the same equations of motion for the gauge invariant quantities, the arbitrariness of the gauge dependent variables will be destroyed.

If the added variable and its corresponding momentum are second class in comparison to other constraints, they can be eliminated at the end of the gauging process, using their constrained equation. This is a part of calculating procedure of Dirac's  bracket, in which those constraints are strongly applied in the Hamiltonian \cite{3}. 
\section{Symplectic formalism for gauge fixed model}
To start with, we use the first iterated Lagrangian, $ L^{(0)} $,as
\begin{equation}
L^{(0)}=\dot{x}_{\mu} v^{\mu} -\frac{1}{2}e(v_{\mu} v^{\mu}-m^{2})+\dot{v}_{\mu} \theta^{\mu \nu} v_{\nu}+\frac{1}{2}\dot{y}^{2}-y \dot{e}.
\end{equation}
So, the canonical Hamiltonian will be read off as,
\begin{align}\label{H1}
& H_{c}= \dot{x}_{\mu}p^{\mu}+\dot{v}_{\mu}\wp^{\mu}+\dot{e}\Pi+ \dot{y}\pi -L^{(0)} , \nonumber \\
& \quad = \frac{1}{2} \dot{y}^{2}+\frac{1}{2}e(v_{\mu} v^{\mu}-m^{2}), \nonumber \\
& \quad = \frac{1}{2} \pi^{2}+\frac{1}{2}e(v_{\mu} v^{\mu}-m^{2})=V^{(0)}.
\end{align}
Comparing \eqref{H1} with \eqref{0H1}, we see that an extra kinetic term for a particle with unit mass is added to the Hamiltonian, which in the symplectic formalism is expressed as the potential term.
Obtaining the corresponding constraints,
\begin{align}
& \phi_{1\mu}=p_{\mu}-v_{\mu}, \nonumber \\
& \phi_{2\mu}=\wp_{\mu}-\theta_{\mu \nu} v^{\nu}, \nonumber \\
& \phi_{3}=\Pi +y.
\end{align}
We construct the first iterated Lagrangian,
\begin{equation}
\frac{1}{2} \pi^{2}+\frac{1}{2} (v_{\mu} v^{\mu}-m^{2}) -\sum_{i=1}^{3}\dot{\lambda}_{i}\phi^{i} = \dot{x}_{\mu}p^{\mu}+\dot{v}_{\mu}\wp^{\mu}+\dot{e}\Pi+ \dot{y}\pi -L^{(1)}.
\end{equation}
Thus, we have,
\begin{align}\label{L(1)}
& L^{(1)}=\dot{x}_{\mu}p^{\mu}+\dot{v}_{\mu}\wp^{\mu}+\dot{e}\Pi+ \dot{y}\pi -V^{(1)}+\dot{\lambda}_{1\mu}(p_{\mu}-v_{\mu})\nonumber \\
& \qquad +\dot{\lambda}_{2\mu}(\wp_{\mu}-\theta_{\mu \nu} v^{\nu})+\dot{\lambda}_{3}(\Pi +y),
\end{align}
in which $ V^{(1)} $ is the same as $ V^{(0)} $ in \eqref{H1}.

The first iterated symplectic variables and one form are now defined as,
\begin{align}
& \xi_{\bar \alpha}^{(1)}=(x_{\mu},v_{\mu},e,y,p_{\mu},\wp_{\mu},\Pi,\pi,\lambda_{1\mu},\lambda_{2\mu},\lambda_{3}) \nonumber \\
&  \mathcal A_{\bar \beta}^{(1)}=(p_{\nu},\wp_{\nu},\Pi,\pi,\textbf{0}_{\nu},\textbf{0}_{\nu },0 ,0,\phi_{1\nu},\phi_{2\nu},\phi_{3}).
\end{align}
Here, $\bar \alpha $ and $ \bar\beta $ are counted from 1 to 23. 

Now, we obtain the corresponding two form tensor, $\bar{f}_{\alpha \beta}^{(1)}$.

\begin{eqnarray}\label{fbar(1)}
\bar{f}_{\bar \alpha \bar\beta}^{(1)}=
\begin{pmatrix}
   0_{\mu\nu} & -\delta_{\mu\nu} & u_{i \nu}^{T} \\
   \delta_{\mu\nu} & 0_{\mu\nu} & 0_{\mu j} \\
  -u_{i \nu} & 0_{i \nu} & 0_{ij} \\
\end{pmatrix},
\end{eqnarray}
which;
\begin{eqnarray}\label{u}
  u_{i \mu}=\frac{\partial \phi_{i}}{\partial q_{\mu}}.
\end{eqnarray}

We see that \eqref{fbar(1)} is a singular tensor. Hence, it has the following zero mode, which generates one extra secondary constraint.
\begin{align}
n_{\bar\alpha}^{(1)}=\{\textbf{0}_{\mu}, \textbf{0}_{\mu}, -1, 0, \textbf{0}_{\mu}, \textbf{0}_{\mu}, 0, 2, \textbf{0}_{\mu}, \textbf{0}_{\mu}, 1\}.
\end{align}

In order to generating new constraints, we use,
\begin{equation}\label{cons}
\phi_{i}=n_{\alpha}^{(1)}\frac{\partial V^{(1)}}{\partial \xi_{\alpha}^{(1)}}.
\end{equation}
Substituting the first iterative potential \eqref{H1} into \eqref{cons}, we obtain the following secondary constraint.
\begin{equation}
\phi_{4}=2\pi -\frac{1}{2}(v_{\mu}v^{\mu}-m^{2}),
\end{equation}
which is also a scalar. This constraint is the correction of the constraint that imposes the condition of being on the mass shell on the particle.

Thus the  second iterated Lagrangian will be
\begin{align}
& L^{(2)}=\dot{x}_{\mu}p^{\mu}+\dot{v}_{\mu}\wp^{\mu}+\dot{e}\Pi+ \dot{y}\pi +\dot{\lambda}_{1}(p_{\mu}-v_{\mu})+\dot{\lambda}_{2}(\wp_{\mu}-\theta_{\mu \nu} v^{\nu}) \nonumber \\
& \qquad +\dot{\lambda}_{3}(\Pi +y) +\dot{\lambda}_{4}(2\pi -\frac{1}{2}(v_{\mu}v^{\mu}-m^{2})) -V^{(1)}.
\end{align}
Now, the symplectic variables and one form extend to the following form,
\begin{align}
& \xi_{\bar \alpha}^{(2)}=(x_{\mu},v_{\mu},e,y,p_{\mu},\wp_{\mu},\Pi,\pi,\lambda_{1\mu},\lambda_{2\mu},\lambda_{3},\lambda_{4}) \nonumber \\
& \mathcal A_{\bar \beta}^{(2)}=(p_{\nu},\wp_{\nu},\Pi,\pi,\textbf{0}_{\nu},\textbf{0}_{\nu},0 ,0,\phi_{1\nu},\phi_{2\nu},\phi_{3},\phi_{4}) ,
\end{align}
where, $ \bar \alpha $ and $ \bar \beta $ are counted now from $ 1 $ to $ 24 $.

Calculating the components of $\bar{f}_{\bar\alpha \bar\beta}^{(2)}$, we will have,
\begin{eqnarray}\label{fbar(2)}
\bar{f}_{\bar\alpha \bar\beta}^{(2)}=
\begin{pmatrix}
   0_{\mu\nu} & -\delta_{\mu\nu} & u_{i \nu}^{T} & v_{i' \nu}^{T} \\
   \delta_{\mu\nu} & 0_{\mu\nu} & 0_{i \nu} & w_{i' \nu}^{T} \\
  -u_{i \nu} & 0_{\mu j} & 0_{ij} & 0_{ij'} \\
  -v_{i' \nu} & -w_{i' \nu}& 0_{i'j} & 0_{i'j'} \\
\end{pmatrix},
\end{eqnarray}
in which, $v_{\alpha}$ and $w_{\alpha}$ are defined as fallow,
\begin{eqnarray}\label{vwbar}
\nonumber v_{i'\mu }=\frac{\partial  \phi_{i'}}{\partial q^{\mu}},  \\
w_{i'\mu }=\frac{\partial \phi_{i'}}{\partial p^{\mu}}.
\end{eqnarray}

We see that \eqref{fbar(2)} is non-singular. The inverse of $f_{\alpha \beta}^{(2)}$ gives the usual Dirac brackets among the physical variables.

Now, it is the time to start symplectic embedding process. In this stage the original phase space of the theory will be expanded by adding an extra function $G$, depending on the original phase space and the WZ variable $ \sigma $ .

\begin{equation}
G(x_{\mu},v_{\mu},e,y,p_{\mu},\wp_{\mu},\Pi , \pi ,\lambda_{i}, \sigma)=\sum^{\infty}_{n=0} \mathcal{G}  ^{(n)}.
\end{equation}
The generator of $ L_{WZ} $ satisfies the following boundary condition,
\begin{equation}
G(x_{\mu},v_{\mu},e,y,p_{\mu},\wp_{\mu},\Pi , \pi ,\lambda_{i},\sigma=0)= \mathcal{G}^{(0)}=0.
\end{equation}

Introducing the new term, G, into the Lagrangian \eqref{L(1)}, we will have,
\begin{align}\label{L(1)bar}
& \tilde{L}^{(1)}=L^{(1)}+L_{WZ} , \nonumber \\
& \qquad =L^{(1)}+G(x_{\mu},v_{\mu},e,y,p_{\mu},\wp_{\mu},\Pi , \pi , \sigma).
\end{align}
The symplectic variables and one form can be read off as,
\begin{align}\label{var sym}
& \tilde{\xi}_{\bar \alpha}^{(1)}=(x_{\mu},v_{\mu },e,y,p_{\mu},\wp_{\mu},\Pi,\pi,\lambda_{1\mu},\lambda_{2\mu},\lambda_{3},\sigma), \nonumber \\
&  \tilde{\mathcal A}_{\bar \beta}^{(1)}=(p_{\n},\wp_{\nu},\Pi,\pi,\textbf{0}_{\nu},\textbf{0}_{\nu},0,0,\phi_{1\nu},\phi_{2\nu},\phi_{3},0). 
\end{align}
Now, $ \bar \alpha $ and $ \bar \beta $ are counted now from $ 1 $ to $ 25 $. Then, we can compute the symplectic two form $\tilde{f}^{(1)}_{\bar\alpha \bar\beta}$,
\begin{eqnarray}\label{f1bar}
 \tilde{f}^{(1)}_{\bar \alpha\bar\beta}= \begin{pmatrix}
 \bar f_{\bar\alpha \bar\beta}^{(1)} &  0_{\bar\alpha 1}\\
0_{1\bar\beta} & 0_{1\times1} \\
\end{pmatrix} .
\end{eqnarray}
Desirably, this tensor is obviously singular. Consequently, it has two independent null vectors as follows,
\begin{align}
& \tilde{ n}_{1\bar\alpha}^{(1)}=\{\textbf{0}, \textbf{0}, 0, 0, \textbf{0}, \textbf{0}, 0, 0, \textbf{0}, \textbf{0}, 0, 1 \}, \nonumber \\
&   \tilde{ n}_{2\bar\alpha}^{(1)}=\{\textbf{0}, \textbf{0}, -1, 0, \textbf{0}, \textbf{0} , 0, 2, \textbf{0},\textbf{0}, 1, 0\}.
\end{align}
We combine $\tilde{n}_{\bar \alpha}$ as the linear combination of the corresponding null vectors.
\begin{align}\label{n alpha}
& \tilde{n}_{\bar \alpha}=\sum_{i}\tilde{n}^{(1)}_{i \bar \alpha}\nonumber \\
&\qquad = \begin{matrix}
 (n^{(1)}_{\alpha} & a ) \\
 \end{matrix},
\end{align}

where, $ a $ is a constant coefficient which makes a linear combination of null vectors. It goes without saying that no more constraint is generated using these null vectors $ \tilde{n}_{\bar\alpha} $. Thus, the first order correction term in $ \sigma $, i.e. $ \mathcal{G}^{(1)} $ , is determined after an integration process,
\begin{equation}
\tilde{n}_{\bar\alpha}\frac{\partial  \tilde{V}^{(1)}}{\partial \tilde{\xi}_{\bar\alpha}^{(1)}}=\frac{\partial \mathcal{G}^{(1)}}{\partial \sigma}.
\end{equation}
Using the  first iterative potential \eqref{H1}, we will have,
\begin{equation}
\tilde{n}_{\bar\alpha}\frac{\partial  \tilde{V}^{(1)}}{\partial \tilde{\xi}_{\bar\alpha}^{(1)}}=2\pi -\frac{1}{2}(v_{\mu}v^{\mu}-m^{2}).
\end{equation}
After integration we obtain,
\begin{equation}
\mathcal{G}^{(1)}=[2\pi-\frac{1}{2}(v_{\mu}v^{\mu}-m^{2})]\sigma.
\end{equation}
Bringing this result into the first order Lagrangian \eqref{L(1)bar},
\begin{align}
& \tilde{L}^{(1)}_{incomplete}=\dot{x}_{\mu}p^{\mu}+\dot{v}_{\mu}\wp^{\mu}+\dot{e}\Pi+ \dot{y}\pi +\sum_{i=1}^{3}\dot{\lambda}_{i}\phi^{i} \nonumber \\
& \qquad \quad-\frac{1}{2}\pi^{2}-\frac{1}{2} e(v_{\mu}v^{\mu}-m^{2})+[2\pi-\frac{1}{2}(v_{\mu}v^{\mu}-m^{2})]\sigma.
\end{align}
Now, we rewrite the $ \tilde{V}^{(1)} $ in the following form,
\begin{equation}
 \tilde{V}^{(1)}=\frac{1}{2}\pi^{2}+\frac{1}{2}e(v_{\mu}v^{\mu}-m^{2})-[2\pi-\frac{1}{2}(v_{\mu}v^{\mu}-m^{2})]\sigma.
\end{equation}
So,
\begin{equation}
\tilde{n}_{\bar\alpha}\frac{\partial  \tilde{V}^{(1)}}{\partial \tilde{\xi}_{\bar\alpha}^{(1)}}=-4\sigma.
\end{equation}
After an integration process, the second order correction term in $ \sigma $, will be determined as,
\begin{equation}
\mathcal{G}^{(2)}=-2\sigma^{2}
\end{equation}
Introducing the above result into the \eqref{L(1)bar}, we get the desired Lagrangian,
\begin{align}\label{L com}
& \tilde{L}^{(1)}=\dot{x}_{\mu}p^{\mu}+\dot{v}_{\mu}\wp^{\mu}+\dot{e}\Pi+ \dot{y}\pi +\sum_{i=1}^{3}\dot{\rho}_{i}\phi^{i}-\frac{1}{2}\pi^{2} \nonumber \\
& \qquad \qquad -\frac{1}{2}(v_{\mu}v^{\mu}-m^{2})+[2\pi-\frac{1}{2}(v_{\mu}v^{\mu}-m^{2})]\sigma-2 \sigma^{2}.
\end{align}
We see that in the added terms to the Lagrangian, there is nothing about the NC parameter. This shows that the terms added to the spatial part of configuration space is not noncommutative any more, i.e. it is commutative.

Rewriting the $  \tilde{V}^{(1)} $, we will have,
\begin{equation}
 \tilde{V}^{(1)}=V^{(0)}-[2\pi-\frac{1}{2}(v_{\mu}v^{\mu}-m^{2})]\sigma+2 \sigma^{2}.
\end{equation}
Using zero mode \eqref{n alpha}, we obtain,
\begin{align}
& \tilde{n}_{\bar\alpha}\frac{\partial  \tilde{V}^{(1)}}{\partial \tilde{\xi}_{\alpha}^{(1)}}=0.
\end{align}
The zero mode $ \tilde{n}_{\bar\alpha} $ does not produce a new constraint. Thus, this model has a symmetry and it is the generator of an infinitesimal gauge transformation. Therefore, all correction terms $ \mathcal{G}^{(n)} $ with $n\geq 3$ vanish.

Now, it is about time we obtain the invariant canonical Hamiltonian.

\begin{equation}
\tilde{L}^{(0)}=\dot{x}_{\mu}p^{\mu}+\dot{v}_{\mu}\wp^{\mu}+\dot{e}\Pi+ \dot{y}\pi +\lambda_{i}\phi^{i}  -\tilde{V}^{(1)}.
\end{equation}
Thus,
\begin{align*}
& H_{c}=\tilde{V}^{(1)}-\lambda_{i}\phi^{i} .
\end{align*}

\subsection{Gauge transformation generators of the gauged model}

In order to obtain gauge symmetries of the model, one can use  the Poisson brackets of the first class constraints and symplectic variables\eqref{var sym}  via the following relation \cite{Shirzad, Henneaux2},
\begin{equation}\label{Shirzad henneaux}
 \delta \tilde{\xi}^{(1)}_{\bar \alpha}=\{\tilde{\xi}^{(1)}_{\bar \alpha},\phi_{j}\}\epsilon_{j}.
\end{equation}

Also, It has been shown that the null vector \eqref{n alpha} are the generators of gauge transformations of the symplectic variables \eqref{var sym} \cite{8,kim}.

\begin{align}\label{Gauge transf}
\begin{array}{ll}
  \delta x_{\mu}=0, 	 & \qquad \delta p_{\mu}=0, \\
  \delta v_{\mu}=0, & \qquad \delta \wp_{\mu}= 0, \\
  \delta e=-\epsilon_{1}, & \qquad \delta \Pi=0,  \\
  \delta y=0, & \qquad \delta \pi=0,  \\ 
  \delta \lambda_{1\mu}=0,\\
 \delta \lambda_{2\mu}=0,\\
  \delta \lambda_{3}=\epsilon_{1},\\
  \delta\sigma=\epsilon_{2}.
\end{array}
\end{align}

We can see that the results obtained from \eqref{Shirzad henneaux} is the same as the infinitesimal gauge transformations \eqref{Gauge transf} obtained by using the null vector \eqref{n alpha}. Apparently, the Lagrangian \eqref{L com} is invariant under the transformations \eqref{Gauge transf}.

\section{Poisson structure of the gauged model}

Calculating all constraints corresponding momenta,
\begin{align}
& p_{\lambda_{1\mu}}=\frac{\partial \tilde{L}^{(0)}}{\partial \dot{\lambda}^{1\mu}}  : \rightarrow  \tilde{\phi}_{1\mu} = p_{\lambda_{1\mu}} , \nonumber \\
& p_{\lambda_{2\mu}}=\frac{\partial \tilde{L}^{(0)}}{\partial \dot{\lambda}^{2\mu}}  : \rightarrow \tilde{\phi}_{2\mu} = p_{\lambda_{2\mu}} , \nonumber \\
& p_{\lambda_{3}}=\frac{\partial \tilde{L}^{(0)}}{\partial \dot{\lambda}^{3}}  : \rightarrow \tilde{\phi}_{3} = p_{\lambda_{3}} , \nonumber \\
& p_{\sigma}=\frac{\partial \tilde{L}^{(0)}}{\partial \dot{\sigma}}  : \rightarrow \tilde{\phi}_{4} =  p_{\sigma} .
\end{align}

Now, we check out consistency conditions. The total Hamiltonian defined by adding primary constraints to canonical Hamiltonian as,
\begin{equation}
\tilde{H}_{T}=\tilde{H}_{c}+\lambda^{i}\tilde{\phi}_{i}.
\end{equation}
Also,
\begin{align}
& 0=\dot { \tilde{\phi}}_{i}=\{ \tilde{\phi}_{i},\tilde{H}_{T} \}, \nonumber \\
& 0= \{ \tilde{\phi}_{i},\tilde{H_{c}} \}+\lambda^{j}\{ \tilde{\phi}_{i},\tilde{\phi_{j}} \}.
\end{align}
We know that $ \psi_{i}=\{ \tilde{\phi}_{i},\tilde{H_{c}} \} $ and also $ \{ \tilde{\phi}_{i},\tilde{\phi_{j}} \}=0 $, thus,
\begin{equation}
\psi_{i}=0,
\end{equation}
and:
\begin{equation}
0=\dot{\psi_{i}}  =\{ \tilde{\psi}_{i},\tilde{H}_{c} \}+\lambda^{j}\{ \tilde{\phi}_{i},\tilde{\psi_{j}} \}.
\end{equation}

Considering weakly vanishing of $ \{ \tilde{\phi}_{i\mu},\tilde{H}_{T} \} $, we obtain,
\begin{align}
& \{ \tilde{\phi}_{1\mu},\tilde{H}_{T} \} =\psi_{1\mu}, \nonumber \\
& \{ \tilde{\phi}_{2\mu},\tilde{H}_{T} \} =\psi_{2\mu}, \nonumber \\
& \{ \tilde{\phi}_{3},\tilde{H}_{T} \} =\psi_{3} , \nonumber \\
& \{ \tilde{\phi}_{4},\tilde{H}_{T} \} =\psi_{4} . 
\end{align}

Now, calculating the consistency condition of $ \psi_{i}s$, we obtain the complete chain of constraints as,
\begin{align}
& \tilde{\phi}_{1\mu} \rightarrow \psi_{1\mu} \rightarrow \Lambda_{1\mu} \nonumber \\
& \tilde{\phi}_{2\mu} \rightarrow \psi_{2\mu} \rightarrow \Lambda_{2\mu} \nonumber \\
& \tilde{\phi}_{3} \rightarrow \psi_{3} \rightarrow \Lambda_{3} \nonumber \\
& \tilde{\phi}_{4} \rightarrow \psi_{4}  
\end{align}

Computing all corresponding Poisson brackets matrix's components, we obtain the following diagram.\\
\begin{table}[htp]
\begin{center}
\begin{tabular}{|c|c|c|c|c|c|c|c|c|c|c|c|}
  \hline

 & $\tilde{\phi}_{1\nu}$ & $\tilde{\phi}_{2\nu}$ & $  \tilde{\phi}_{3} $ & $  \tilde{\phi}_{4} $ & $ \psi_{1\nu} $ & $  \psi_{2\nu} $ & $ \psi_{3} $& $  \psi_{4} $ & $  \Lambda_{1\nu} $ & $  \Lambda_{2\nu} $& $  \Lambda_{3} $\\ \hline
  
   $\tilde{\phi}_{1\mu}$ & \textbf 0 & \textbf 0 &0 &0 &\textbf 0 &\textbf 0 &\textbf 0 &\textbf 0 &\textbf 0 & $ - \textbf{1} $ &0  \\ \hline
   
  $\tilde{\phi}_{2\mu}$ & \textbf 0 & \textbf 0 &0 &0 &\textbf 0 &\textbf 0 &\textbf 0 &\textbf 0 &- \textbf 1 & $ -2\theta_{\mu\nu} $ &0 \\ \hline
  
  $  \tilde{\phi}_{3} $ & 0 & 0 &0 &0 &0 &0 &0 & 0 & 0 & 0 & 0 \\ \hline
  
  $  \tilde{\phi}_{4} $ & 0 & 0 &0 &0 &0 &0 &0 & -4 & 0 & $ v_{\mu} $ & 2 \\ \hline
  
 $ \psi_{1\mu} $ & \textbf 0 & \textbf 0 &0 &0 &\textbf 0 & -\textbf 1 &\textbf 0 & \textbf 0 & \textbf 0 & \textbf 0 & 0 \\ \hline
 
 $  \psi_{2\mu} $  & \textbf 0 & \textbf 0 &0 &0 & \textbf 1 & \textbf 0 &\textbf 0 & $  -v_{\mu}$ & \textbf 0 & $ (e+\sigma)\delta_{\mu\nu} $ & $ v_{\mu} $ \\ \hline
 
  $ \psi_{3} $ & \textbf 0 & \textbf 0 &0 &0 &\textbf 1 & \textbf 0 &\textbf 0 & -\textbf 2 & \textbf 0 & $ v_{\mu} $ & 1 \\ \hline
  
  $  \psi_{4}$ &  0 &  0 &0 &4 &  0 & $ v_{\mu} $ &2 & 0 & 0 & 0 & 0 \\ \hline
  
  $  \Lambda_{1\mu} $& \textbf 0 & \textbf 1 &0 &0 & \textbf 0 & \textbf 0  &\textbf 0 & \textbf 0 & \textbf 0 & \textbf 0 & 0 \\ \hline
  
 $  \Lambda_{2\mu} $ & \textbf 1 & $  -2\theta_{\mu\nu}$ &0 &$  -v_{\mu}$ & \textbf 0 & $ -(e+\sigma)\delta_{\mu\nu} $  &$  -v_{\mu}$ &\textbf 0 &\textbf 0 &\textbf 0 & 0\\ \hline
 
 $  \Lambda_{3} $ & 0 & 0 &0 &-2 & 0 & $ -v_{\mu}$  &-1 & 0 & 0 & 0 & 0\\ \hline
\end{tabular}
\caption{Poisson brackets between extended phase space variables of the gauged model}
\end{center}
\end{table}

It is evident that $ \tilde{\phi}_{3} $ is a first class constraint. Eliminating this constraint and calculating the determinant of the Poisson bracket matrix, we get zero, which shows that there are more first class constraints between them.

\subsection{Purifying the constraint structure}
In order to find the Poisson structure of a second class system, Dirac presented a new version of Poisson bracket, because in any general case, there is no way to calculate some phase space coordinates in the form of others. This is due to the fact that second class constraints are not necessarily in the form of linear functions of phase space coordinates.

It goes without saying that in the set of obtained constraints, second class couples $ (\lambda_{2\mu},p_{\lambda_{2\mu}}) $ and $ (y,\pi) $, which are linearly appeared in constraints equations. Using these equations, one can impose them directly in the Hamiltonian, without calculating Dirac's brackets. Because, these couples are just appeared in four constraint equations, there will be no ambiguity to solve them from those equations.

So, we will have,
\begin{align}
&  \tilde{\phi}_{1\mu} = p_{\lambda_{1\mu}}, \nonumber \\
&  \tilde{\phi}_{4} =  p_{\sigma}  ,\nonumber \\
&  \psi_{1\mu}= p_{\mu}-v_{\mu}  , \nonumber \\
& \psi_{2\mu} =  \wp_{\mu}-\theta_{\mu \nu}v^{\nu} ,\nonumber \\
& \psi_{4}=4\sigma-\frac{1}{2}(v_{\mu}v^{\mu}-m^{2})  ,\nonumber \\
& \Lambda_{2\mu}=-v_{\mu}(e+\sigma)-\lambda_{1\mu}  ,\nonumber \\
& \Lambda_{3}=-2\sigma-\frac{1}{2}(v_{\mu}v^{\mu}-m^{2}) .
\end{align}

Now, solving $ \sigma $ from the last equation and putting the result in $ \psi_{4} $, we obtain,
\begin{align}\label{rcons1}
& \psi_{4}= v_{\mu}v^{\mu}-m^{2}   ,\nonumber \\
& \Lambda_{2\mu}=-v_{\mu}(e+\sigma)-\lambda_{1\mu}  ,\nonumber \\
& \Lambda_{3}=\sigma .
\end{align}

Again, we eliminate the phase space coordinates which are appeared linearly in \eqref{rcons1}. Projecting constraints on the surface of others, we find their pure content. So, we easily find first class and second class constraints.

Redefining all remained constraints,
\begin{align}\label{rcons2}
&  \Psi^{(1)}_{\mu}= p_{\mu}-v_{\mu}  , \nonumber \\
& \Psi^{(2)}_{\mu} =  \wp_{\mu}-\theta_{\mu \nu}p^{\nu} ,\nonumber \\
& \Phi^{(1)}=p_{\mu}p^{\mu}-m^{2} ,\nonumber \\
& \Phi^{(2)}=p_{\lambda 3}.
\end{align}
Thus, we could find a second class constraint couple $  (\Psi^{(1)}_{\mu}, \Psi^{(2)}_{\mu}) $ and a scalar first class one, $ (\Phi^{(1)},\Phi^{(2)}) $. We notice that $  \Phi^{(1)}$ is related to the mass shell gauge symmetry.

Considering these modifications, one can read off the reduced canonical Hamiltonian as,
\begin{equation}
\tilde H_{c}^{red}=-4\sigma^{2}-\lambda_{1\mu}(p^{\mu}-v^{\mu})-\lambda_{2\mu}(\wp^{\mu}-\theta^{\mu\nu}p_{\nu}).
\end{equation}

Calculating Dirac's bracket shows that our canonical structure will not change.

\subsection{Quantization of the primary and the gauged model}

Taking into the account two primary constraints of the original model, $ \phi_{1} $ and $ \phi_{2} $, and two second class constraints, $ \Psi^{(1)} $ and $ \Psi^{(2)} $ of the gauged model, and calculating their corresponding Poisson  brackets matrix, we will have,
 \begin{equation} \label{deltaij}
 \Delta_{ij}=\left(
  \begin{array}{cc}
    0 & -\textbf{I} \\
   \textbf{I} & 0  \\
  \end{array}
\right).
 \end{equation}

In order to determine all Dirac brackets of the original and gauged model, we put the inverse of $ \Delta_{ij} $, in the following formula,
\begin{equation}\label{Dirac brackets}
 \{\xi_{\bar\alpha},\xi_{\bar\beta}\}^{\ast}= \{\xi_{\bar\alpha},\xi_{\bar\beta}\}-\{\xi_{\bar\alpha},\Psi^{(i)}\} \Delta_{ij}^{-1}\{\Psi^{(j)},\xi_{\bar\beta}\}.
\end{equation}

\begin{table}[h]
\begin{center}
\begin{tabular}{|c|c|c|}
  \hline
 $ \{,\}^{\ast} $  &   Primary Model &  Gauged Model \\   \hline
   $(x_{\mu},x_{\nu})$ & $ 2\theta_{\mu\nu}$ & $ 2\theta_{\mu\nu}$ \\ \hline
  $(x_{\mu},p_{\nu})$ & $\delta_{\mu\nu}$ & $\delta_{\mu\nu}$ \\ \hline
    $(x_{\mu},v_{\nu})$ & $ \delta_{\mu\nu} $ & $ \delta_{\mu\nu} $ \\ \hline
     $(p_{\mu},\wp_{\nu})$ & $ \delta_{\mu\nu} $ & $ \delta_{\mu\nu} $ \\ \hline
       $(v_{\mu},\wp_{\nu})$ & $ \delta_{\mu\nu} $ & $ \delta_{\mu\nu} $ \\ \hline
  $(e,\Pi)$ & $ 1 $ & $ 1 $ \\ \hline
  $(y,\pi)$ &  $N/A$ & $N/A$ \\ \hline
  $(\lambda_{1\mu},p_{\lambda1\nu})$ & $ \delta_{\mu\nu} $ & $ \delta_{\mu\nu} $ \\ \hline
   $ (\lambda_{2\mu},p_{\lambda2\nu}) $ & $\delta_{\mu\nu} $ & $ \delta_{\mu\nu} $ \\ \hline
  $(\lambda_{3},p_{\lambda3})$ & $ N/A $ & $ 1 $ \\ \hline
  $(\sigma,p_{\sigma})$ & $ N/A $ & $ 1 $ \\ \hline
\end{tabular}
\caption{Dirac brackets between extended phase space variables}\label{tab2}
\end{center}
\end{table}

By characterizing first class constraints and Dirac brackets of a classical system, the Hilbert space of the quantum states of the model will be fully available. Thus, according to Dirac prescription one can obtain the quantized version of the model. 
\begin{equation}
\{A,B\}^{\ast} \rightarrow \frac{1}{i \hbar}[A,B], \qquad \hat \phi_{i}\mid phys> =0.
\end{equation}

Here, $\hat\phi_{i}$ is a quantized version of first class constraints. Having two second class constraints of \eqref{rcons2}, which transform fundamental commutators to the commutators of table \ref{tab2}, and two first class constraints of \eqref{rcons2} in their quantum form, we get two more quantum equations than the Schrodinger equation in quantized version.

Also, according to the content of table \ref{tab2}, one can see that in the quantized model, we derive a NC structure in the momentum part of the phase space, despite the intact NC intrinsic structure of coordinates. 

\section{Conclusion}
In this article we gauge the relativistic particle model on a NC surface, using the symplectic formalism of the constrained plane. We introduce a short cut to construct the gauged Lagrangian without calculating the whole symplectic procedure. We also show that although this formalism is more reliable than its congeners, we see that it may be defeated in some cases which the production of secondary constraints is failed. In order to propose a solution  we show that adding an auxiliary field to impose more second class constraints into the system is remedial. In continue, we illustrate that the added extra field should be eliminated, using the equation of constraint directly after gauging is applied. We also extracted the corresponding gauge transformation generators of the gauged model. At the end, comparing the Poisson structure of the gauged and ungauged model, we see that noncommutativity remains intact after gauging is applied.

\end{document}